\documentclass[reprint,showpacs,amsmath,amssymb,floatfix,apl,superscriptaddress,showkeys,aip]{revtex4-1}

\usepackage{graphicx}
\usepackage{epsfig}
\usepackage{dcolumn}
\usepackage{bm}
\usepackage{times}
\usepackage{color}
\usepackage{soul}
\begin{document}


\title{Eigenvalue bounds in one dimensional Schr\"odinger's equation with ultra-short potentials}

\author{Gabriel Gonz\'alez}
\email{gabriel.gonzalez@uaslp.mx}
\affiliation{C\'atedras Conacyt, Universidad Aut\'onoma de San Luis Potos\'i, San Luis Potos\'i, 78000 MEXICO}
\affiliation{Coordinaci\'on para la Innovaci\'on y la Aplicaci\'on de la Ciencia y la Tecnolog\'ia, Universidad Aut\'onoma de San Luis Potos\'i,San Luis Potos\'i, 78000 MEXICO}

\date{\today}

\begin{abstract}
The problem of a particle localized in a ultra-short potential in one dimension is considered. By proposing a general solution to Schr\"odinger's equation we show that the energy spectra and the probability of the particle have definite bounds for an arbitrary ultra-short potential. These results are relevant for the confinement of particles in nanodevices.
\end{abstract}


\maketitle

\section{Introduction}
One of the basic problems in quantum mechanics is to find the energy eigenvalues of a microsystem with a given.\cite{d2} Unfortunately, there are only a few potentials for which the Schr\"odinger equation is exactly solvable and there is no systematic procedure for dealing with even the simpler problems in one-dimensional quantum mechanics.\cite{lau} Now numerical methods usually allow accurate solutions for most problems.\cite{ma} However, the profile of the potential confining the particle in a microsystem only in few cases is known with sufficient precision. Approximate potentials or effective potentials are often very useful for the numerical and theoretical calculations.\cite{merz} However, the knowledge of the realistic profile of the confinement potential is necessary to obtain the accurate physical properties of the microsystem.\cite{shan}\\
With the rapid progress in nanoscale fabrication, nanodevices in which the particle is confined in one or more directions have been successfully fabricated.\cite{harr}
The confinement potential in which the electron's wave function is comparable with the size of the particle ranges from 1 to 25 nm. As a result of these spatial constraints the localized particle responds by adjusting its energy. The problem that I address in this communication is whether we can extract physical information when a particle is localized in a ultra-short confinement potential without knowing the explicit form of the realistic potential.
\section{Approximate Method}
Consider a particle of mass $m$ that moves in one dimension under the following ultra-short potential
\begin{equation}
U(x)=\left\{ \begin{array}{ll}
																				f(x) & \mbox{if $0<x<\delta x$}\\
																				0 & \mbox{otherwise},
																				\end{array}\right.
\label{eq1}
\end{equation} 
where $\delta x$ is the confining dimension where we want to localized the particle. Since we expect to have bound states in the ultra-short potential given by eq. (\ref{eq1}) we propose the following general solution for the time-independent Schr\"odinger equation
\begin{equation}
\psi(x)=\left\{ \begin{array}{ll}
																				Ae^{kx} & \mbox{if $x<0$}\\
																				\psi_1(x) & \mbox{if $0<x<\delta x$}\\
																				Be^{-kx} & \mbox{if $x>\delta x$}
																				\end{array}\right.
\label{eq2}
\end{equation}
where $A$ and $B$ are constants, $k=\sqrt{-2mE}/\hbar$ and $\psi_1$ represents the wave function in the region of the ultra-short potential. The next step is to apply the appropiate boundary conditions at $x=0$ and $x=\delta x$ over the wave function and its derivative. In this case, the first set of boundary conditions over $x=0$ read
\begin{eqnarray}
\label{eq3}
A&=&\psi_1(0) \\ \nonumber
kA&=&\psi^{\prime}_1(0)
\end{eqnarray}
Assuming that $\psi_1(0)\neq 0$, we conclude from eq.(\ref{eq3}) that $k=~\psi^{\prime}_1(0)/\psi_1(0)$. The other set of boundary conditions over $x=~\delta x$ read
\begin{eqnarray}
\label{eq4}
\psi_1(\delta x)&=&Be^{-k\delta x} \\ \nonumber
\psi^{\prime}_1(\delta x)&=&-kBe^{-k\delta x}
\end{eqnarray}
From eq. (\ref{eq4}) we conclude that $\psi^{\prime}_1(\delta x)=-k\psi_1(\delta x)$. If we now take into account that $\delta x$ is a really small distance, we can write up to first order of approximation $\psi^{\prime}_1(\delta x)~\approx-k\left(\psi_1(0)+\delta x\psi_1^{\prime}(0)\right)~=-k\psi_1(0)\left(1+k\delta x\right)$. Using this approximation we can write down the general solution of the time-independent Schr\"odinger equation for ultra-short potentials which is given by
\begin{eqnarray}
\psi(x)=\left\{ \begin{array}{ll}
																				\psi_1(0)e^{kx} & \mbox{if $x<0$}\\
																				\psi_1(x) & \mbox{if $0<x<\delta x$}\\
																				\psi_1(0)e^{-k(\delta x-x)}\left(1+k\delta x\right) & \mbox{if $x>\delta x$}
																				\end{array}\right.
\label{eq5}
\end{eqnarray}
We can apply now the normalization condition to eq. (\ref{eq5}) which reads
\begin{equation}
1=\int_{-\infty}^{0}\!\!\!\!\!\!\psi^2_1(0)e^{2kx}dx+\int_{0}^{\delta x}\!\!\!\!\!\!\psi^2_1(x)dx +\int_{\delta x}^{\infty}\!\!\!\!\!\!\psi^2_1(0)e^{2k(\delta x-x)}\left(1+k\delta x\right)^2dx
\label{eq6}
\end{equation}
where we have taken into account in eq.(\ref{eq6}) that the wave function for bound states in one dimensional systems can be chosen to be purely real. \\
The second integral in the right side of eq. (\ref{eq6}) represents the probability of finding the particle between $0$ and $\delta x$, and can be rewritten up to second order of approximation by integrating by parts, which gives
\begin{equation}
\int_{0}^{\delta x}\!\!\!\!\!\!\psi^2_1(x)dx\approx\delta x \psi_1^2(\delta x)-\left(\delta x\right)^2\psi_1(\delta x)\psi_1^{\prime}(\delta x)
\label{eq7}
\end{equation} 
substituting $\psi^{\prime}_1(\delta x)~=-k\psi_1(0)\left(1+k\delta x\right)$ into eq. (\ref{eq7}) we end up with
\begin{equation}
\int_{0}^{\delta x}\!\!\!\!\!\!\psi^2_1(x)dx=\psi_1^2(\delta x)\left[\delta x+k(\delta x)^2\right]
\label{eq8}
\end{equation}
Using the fact that $\psi_1^2(\delta x)\approx\psi_1^2(0)+2\delta x\psi_1(0)\psi_1^{\prime}(0)$ and expressing everything in terms of $\psi_1(0)$ we finally have that the probability of finding the particle in the ultra-short potential up to second order of approximation is given by 
\begin{equation}
\int_{0}^{\delta x}\!\!\!\!\!\!\psi^2_1(x)dx=\delta x\psi_1^2(0)\left(1+3k\delta x\right)
\label{eq9}
\end{equation}
Substituting eq.(\ref{eq9}) into eq.(\ref{eq6}) we have the following equation
\begin{equation}
1=\frac{\psi_1^2(0)}{2k}\left[2+2k\delta x+k^2\left(\delta x\right)^2\right]+\psi_1^2(0)\delta x+3k\left(\delta x\right)^2\psi_1^2(0)
\label{eq10}
\end{equation}
Let $z=k\delta x$ and $P=\psi_1^2(0)\delta x$, then eq.(\ref{eq10}) reads
\begin{equation}
7Pz^2+\left(4P-2\right)z+2P=0
\label{eq11}
\end{equation}
Solving eq.(\ref{eq11}) for $z$ we have
\begin{equation}
z=\frac{1-2P\pm\sqrt{1-4P-10P^2}}{7P}
\label{eq12}
\end{equation}
Since $z$ is a real positive number we have to satisfy the following inequality
\begin{equation}
1-4P-10P^2\geq 0
\label{eq13}
\end{equation}
Solving eq. (\ref{eq13}) and using the fact that $P=\psi^2_1(0)\delta x>0$ we find the following inequality 
\begin{equation}
0<\psi^2_1(0)\delta x\leq\frac{1}{10}\left(\sqrt{14}-2\right)
\label{eq14}
\end{equation}
Equation (\ref{eq14}) is our first important result, which states that the maximum probability of finding the particle in the ultra-short potential up to first order of approximation is $17.4$\%, independently of the profile of the confinement potential. \\ 
Since the higher energy states spend more of their time outside the ultra-short potential we need to choose the negative root in eq.(\ref{eq12}), therefore
\begin{equation}
z=\frac{1-2P-\sqrt{1-4P-10P^2}}{7P}
\label{eq15}
\end{equation}
If we substitute in eq.(\ref{eq15}) the highest probability of finding the particle inside the ultra-short potential, then the upper and lower limits for the energy spectra of the particle will be given by 
\begin{equation}
0<|E|\leq\frac{\hbar^2}{2m\left(\delta x\right)^2}\left(\frac{\sqrt{14}-1}{7-\sqrt{14}}\right)
\label{eq16}
\end{equation}
Equation (\ref{eq16}) is our second important result which states that the energy spectra of the bound state is inversely proportional to the square of the confinement dimension where we want to localized the particle. Of course, we expect eq.(\ref{eq14}) and eq.(\ref{eq16}) to be more accurate for small values of $\delta x$ rather than large values.\\
To illustrate the result given in eq.(\ref{eq16}), consider that we want to trap an electron in a one dimensional ultra-short potential which has a confinement dimension of $\delta x=5$ nm, then we can estimate the range of the energy spectra of the bound state using eq.(\ref{eq16}), which will be given by $0<|E|\leq 1$ meV. 
\section{Conclusions}
In this communication we presented a general solution for Schr\"odinger's equation in one dimension under an arbitrary ultra-short potential. Using the general solution  
we have shown an approximate method which can estimate the energy spectra of the particle localized in an arbitrary ultra-short potential in one dimension. We have demonstrated that the maximum probability of finding the particle in the ultra-short potential is of $17.4$\%. These results are relevant for the confinement of particles in nanodevices.

\section{Acknowledgments}
This work was supported by the program ``C\'atedras CONACYT".\\

\end{document}